\documentclass[pra,twocolumn,amsmath,amssymb,groupaddress,eqsecnum]{revtex4-2}

\usepackage{graphicx,relsize,epstopdf,color,mathtools,bm,newtxtext,newtxmath,braket,rotating,dsfont,mathtools}

\usepackage[hyphenbreaks]{breakurl}
\usepackage[colorlinks=true,linkcolor=blue,citecolor=blue,urlcolor =blue]{hyperref}
\usepackage{float}
\usepackage{flafter}
\usepackage{placeins}
\usepackage{soul}
\usepackage{cancel}
\usepackage{comment}
\newcommand{\Tr}{\mathop{\mathrm{Tr}}\nolimits}

\newcommand{\sinc}{\mathop{\mathrm{sinc}}\nolimits}

\begin{document}

\title{Sampling the spatial coherence of light}

\author{M. Vitek}
\affiliation{Department of Optics, Palack\'y University, 771 46 Olomouc, Czech Republic}

\author{M. Peterek} 
\affiliation{Department of Optics, Palack\'y University, 771 46 Olomouc, Czech Republic}

\author{D. Koutn\'y} 
\affiliation{Department of Optics, Palack\'y University, 771 46 Olomouc, Czech Republic}

\author{M. Pa\'ur}
\affiliation{Department of Optics, Palack\'y University, 771 46 Olomouc, Czech Republic}

\author{L. Motka} 
\affiliation{Department of Optics, Palack\'y University, 771 46 Olomouc, Czech Republic}

\author{B. Stoklasa}
\affiliation{Department of Optics, Palack\'y University, 771 46 Olomouc, Czech Republic}

\author{Z. Hradil} 
\affiliation{Department of Optics, Palack\'y University, 771 46 Olomouc, Czech Republic}

\author{J.~\v{R}eh\'a\v{c}ek} 
\affiliation{Department of Optics, Palack\'y University, 771 46 Olomouc, Czech Republic}

\author{L. L. S\'anchez-Soto}
\affiliation{Departamento de \'Optica, Facultad de F\'{\i}sica, Universidad Complutense, 28040 Madrid, Spain}
\affiliation{Max-Planck-Institut f\"ur die Physik des Lichts, 91058 Erlangen, Germany}
 \affiliation{Institute for Quantum Studies, Chapman University, Orange, CA 92866, USA}

\begin{abstract}
We introduce a novel application of the Hartmann sensor, traditionally designed for wavefront sensing, to measure the coherence properties of optical signals. By drawing an analogy between the coherence matrix and the density matrix of a quantum system, we recast the sensor operation as a quantum estimation problem. We experimentally demonstrate its effectiveness in the regime where  signals from different apertures significantly overlap, enabling information extraction beyond the reach of standard wavefront sensing.
\end{abstract}

\maketitle{}

\section{Introduction} 

Three-dimensional objects radiate complex wavefronts shaped by their intrinsic properties, such as geometry, refractive index, density, and temperature. These wavefronts carry critical  information about the emitter, yet conventional detectors are limited to capturing only two-dimensional intensity and lack sensitivity to phase variations. As a result, wavefront sensing becomes essential for reconstructing the full optical field, enabling a deeper understanding of the object structure and behavior. This capability is crucial in applications such as high-resolution imaging~\cite{Wu:2019aa,Hampson:2021aa}, ophthalmology~\cite{Maeda:2001aa,Dai:2008aa,Vacalebre:2022aa}, optical testing~\cite{Malacara:2007aa}, beam shaping~\cite{Mosk:2012aa,Yu:2015aa} adaptive optics~\cite{Tyson:2022aa,AOpt:2024aa},  and precision metrology~\cite{Rastogi:1997aa}, where extracting phase information provides insights beyond what intensity-based measurements alone can offer.

Wavefront reconstruction can be accomplished through several distinct approaches, each with its own strengths and limitations~\cite{Geary:1995aa}. Broadly, these methods fall into three primary categories: (a) interferometric techniques, which rely on the interference of two beams with a carefully controlled relative phase; (b) methods based on the determination of the wavefront's slope or curvature; and (c) image-based approaches utilizing iterative phase-retrieval algorithms~\cite{Luke:2002aa,Guo:2022aa}. While substantial progress has been made in this area, achieving robust wavefront sensing remains a challenging problem.

Among the various schemes for wavefront reconstruction, the time-honored Shack-Hartmann sensor surely deserves special recognition. The original design incorporated a Hartmann mask, consisting of a regular array of pinholes~\cite{Hartmann:1900aa}. In the late 1960s, Shack revolutionized the design by introducing lens arrays, greatly enhancing the sensor sensitivity~\cite{Platt:2001aa}. With its dynamic range, high optical efficiency, compatibility with white light, and versatility in handling both continuous and pulsed sources, this setup has proven to be an exceptional solution for a variety of applications~\cite{Neal:2002aa}.

The operation of a Hartmann sensor may at first seem intuitively straightforward~\cite{Primot:2003aa}. The wavefront is sampled by an array of pinholes, producing a corresponding set of spots on the recording plane. The local slope of the wavefront at each sample point is inferred from the direction in which most of the light emerges from the respective pinhole—specifically, from the position of the resulting spot. However, this simplified geometrical framework inherently assumes the presence of a well-defined wavefront, implying full coherence of the signal. Furthermore, it requires that the spots remain non-overlapping.

However, emerging applications in fields such as lithography~\cite{Kazazis:2024aa}, optical microscopy~\cite{Booth:2014aa}, and holography~\cite{Peng:21} demand the measurement of signals with increasingly complex temporal and spatial mode structures. In these scenarios, conventional wavefront-sensing strategies prove inadequate, as they are designed to handle spatially coherent signals.

It has been suggested~\cite{Hradil:2010aa, Stoklasa:2014aa} that conventional wavefront sensors may be underrated, as they do not fully exploit the potential of the recorded data. By incorporating tomographic reconstruction techniques, Shack-Hartmann sensors could facilitate the evaluation of the mutual coherence, thus enabling a complete characterization of the signal.

In this paper, we build upon our previous work by incorporating a digital micromirror device (DMD) as an addressable reflective mask in the sensor. Although DMDs are limited to binary amplitude modulation, their rapid and reliable modulation capabilities have enabled a wide range of applications~\cite{Ren:2015aa,Scholes:2019aa}. Recently, they have been explored for characterizing the spatial coherence through interferometric techniques~\cite{Partanen:2014aa, Magalhaes:2021aa, Shirai:21}. However, this approach is limited to assessing the complex degree of coherence, as it relies on measuring a visibility.

We expand the scope of this framework to achieve point-by-point sampling of the mutual coherence. By  leveraging quantum estimation theory~\cite{Helstrom:1976aa}, we surpass the limitations of conventional wavefront sensors, extracting information beyond standard reconstruction methods. This is more than an academic curiosity, as spatial coherence plays a vital role in various applications~\cite{Korotkova:2020aa}.

This paper is organized as follows. In Sec.~\ref{sec:sam} we briefly review the essential ingredients needed to describe partially coherent light and rephrase the action of a Hartmann sensor in a quantum parlance that will prove useful for the remaining discussion. This is applied in Sec.~\ref{sec:Rec}  to devise a feasible DMD-based protocol that allows for the sampling of the mutual intensity. The experimental results are presented in Sec.~\ref{sec:exp}, confirming the reconstruction of the coherence properties of the signal. Finally, our conclusions are summarized in Sec.~\ref{sec:conc}.

\section{Sampling the mutual intensity}
\label{sec:sam}

Partially coherent light cannot be adequately described by intensity or complex amplitude alone. Instead, its properties must be captured through correlation functions, with the second-order correlation  (commonly known as the mutual coherence function) being the most commonly used~\cite{Mandel:1995qy}. To simplify our discussion, we will restrict ourselves to the quasimonochromatic regime, where this correlation reduces to the so-called mutual intensity, 
\begin{equation}
J( \mathbf{x}, \mathbf{x}^{\prime}) = \langle u(\mathbf{x},t) u^{\ast}(\mathbf{x}^{\prime},t) \rangle \, ,
\label{eq:mutint}
\end{equation}
with $u(\mathbf{x},t)$ representing  the complex amplitude. Assuming  $u(\mathbf{x},t)$ is an ergodic random process,  $J$ becomes time independent, allowing the ensemble average $\langle \cdot \rangle$ to be replaced by a time average.

The sensors we are considering here operate by dividing the incoming light wavefront into smaller sections. This is achieved through an array of holes (in a Hartmann sensor) or lenslets (in a Shack-Hartmann sensor), each directing light onto a detector. For a perfectly flat (planar) wavefront, all the  spots align in a regular grid, forming the reference spot field.

When the wavefront is distorted by aberrations, its curvature changes, creating small tilts in different regions. These tilts cause the positions of the image spots to shift. By comparing the shifted positions to their original locations in the reference spot field, we can determine the wavefront shape and any aberrations in the beam. By analyzing the entire pupil, this information is combined into a phase map, which reveals the wavefront structure~\cite{Tyson:2022aa}. Unfortunately, as heralded in the Introduction, when dealing with partially coherent light, this naive picture breaks down, as the concept of a single, well-defined wavefront becomes ambiguous.

In a standard wavefront configuration, the setup is carefully arranged to prevent overlapping spots. This enables the reconstruction of a putative wavefront, but precludes the measurement of spatial coherence. In contrast, we explore an alternative scenario,  where signals are deliberately overlapped. This distinctive setup fosters mutual interference among signals from the subapertures, thereby encoding valuable coherence properties directly into the experimental data.

To facilitate possible generalizations, we frame the following discussion in quantum terms, drawing on the deep mathematical parallels between wave optics and quantum theory. In an obvious Dirac notation, a coherent beam with a complex amplitude $u(\mathbf{x})$ will be assigned to a ket $|u \rangle$, such that $u(\mathbf{x}) = \langle \mathbf{x}| u \rangle$, where $|\mathbf{x} \rangle$ represents a pointlike source located at $\mathbf{x}$. In this framework, partial coherence is naturally described by the density matrix $\varrho$. Consequently,  the mutual intensity \eqref{eq:mutint} can be rewritten as    
\begin{equation}
J(\mathbf{x},\mathbf{x}^{\prime})= \langle \mathbf{x} | \varrho | \mathbf{x}^{\prime} \rangle = \Tr ( \varrho|\mathbf{x} \rangle\langle \mathbf{x}^{\prime}|) \, , 
\end{equation} 
which corresponds to the position representation of the density matrix. 

The key step in our approach is to expand the signal $\varrho$ as a finite superposition of appropriate modes that serve as a computational basis. We denote this basis as $\{ \bm{\psi}_{j} \}$, for $j=1,\ldots,d$, where $d$ represents the effective dimension of the signal. In this basis the mutual intensity appears as a  $d \times d$ nonnegative matrix 
\begin{equation}
J_{ij} = \langle \bm{\psi}_{i} | \varrho | \bm{\psi}_{j} \rangle \, .
\end{equation} 
The choice of basis depends on the specific experimental setup. For instance, in Ref.~\cite{Stoklasa:2014aa} the basis consisted of spatially unbounded modes, which is suitable when there is no spot overlap. In such a scenario, signal components passing through different apertures remain separate, preventing the determination of their mutual coherence.

In contradistinction, when spots overlap, a different election is required. We adopt a basis composed of sharply bounded modes (i.e., modes with vanishing amplitude outside a finite region). While these modes possess an unbounded Fourier spectrum and, consequently, an unlimited range of transverse momenta, they are conveniently characterized in the $x$-representation, which we prioritize in this work.  If, for simplicity, we restrict ourselves to a one-dimensional model, a sensible choice for these modes is 
\begin{equation}
 \psi_{j} (x) = \langle x| \psi_{j} \rangle= \text{rect}  \left ( \frac{x-x_{0}}{a} \right ) e^{i k_{j} x} \, ,
\end{equation}
where the rectangle function describes the action of a single Hartmann aperture centered at $x_{0}$, which is taken as a square aperture of side~$a$.  We assume here that the intensity, phase, and statistical properties evolve slowly relative to the size of the apertures, similar to sampling assumptions in classical wavefront sensors. Additionally, we have limited a single mode (i.e., a single local $\mathbf{k}$ vector) to each aperture. 

The measurement process is described using a positive operator-valued measure (POVM)~\cite{Holevo:2003fv}, represented by the projectors $\Pi_{\xi}$, where $\xi$ corresponds to the specific signal detected at a given pixel on the detector.  Essentially, the POVM provides a mathematical framework for understanding how the measurement extracts information from the incoming light. It encapsulates the operational principles of the DMD, free-space propagation, and detection. In the computational basis, it can be conveniently represented as the measurement matrix 
\begin{equation}
(\Pi_\xi)_{ij} = \psi_{i}^{\prime} (\xi) \; \psi_{j}^{\prime \, \ast} (\xi) \, , 
\end{equation} 
where $\psi_{j}^{\prime} (\xi)$ is the complex amplitude at the detector generated by the $j$th basis mode.

\begin{figure}[t]
\centerline{\includegraphics[width=\columnwidth]{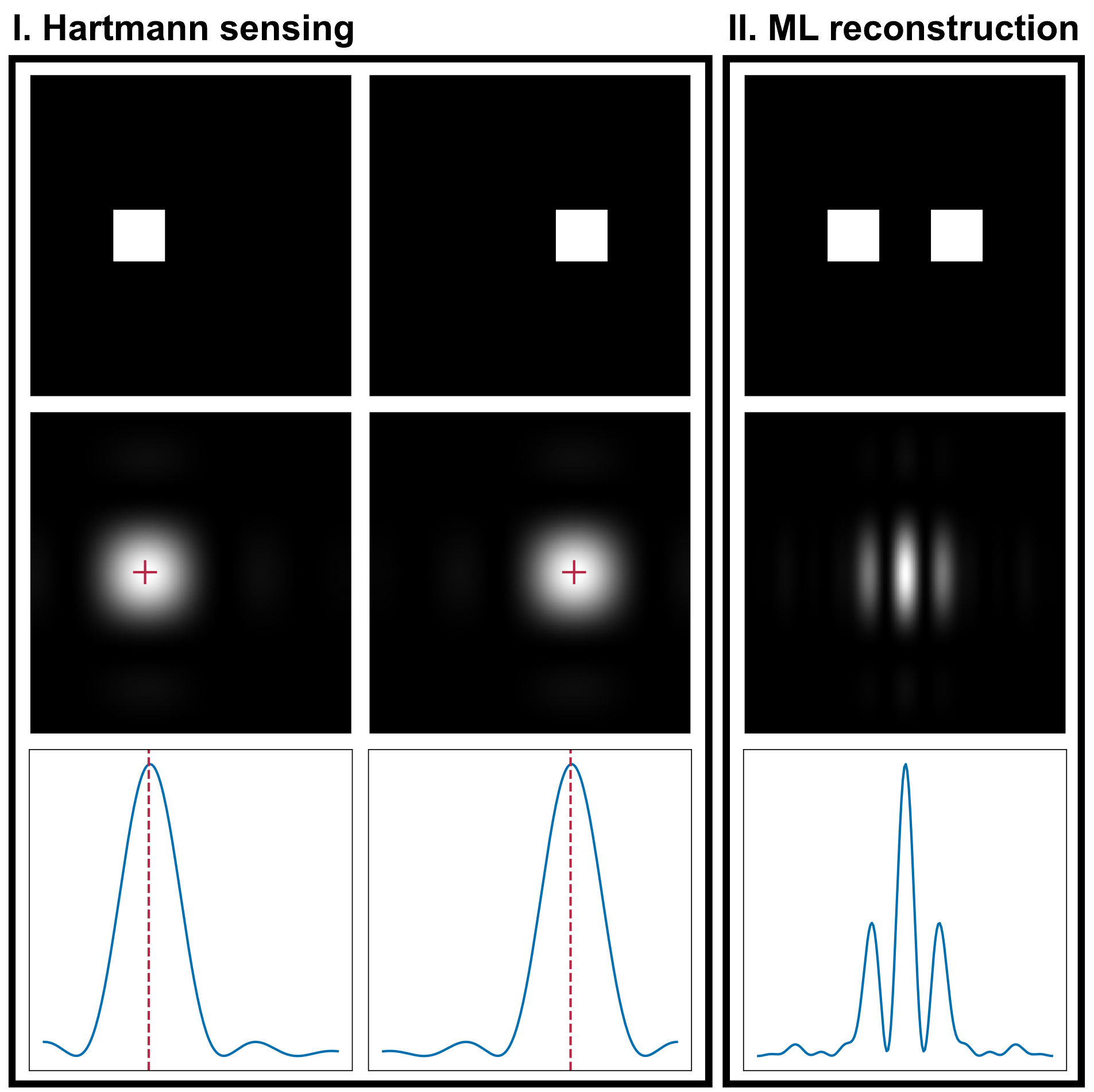}}
\caption{Protocol steps. (Left panel) The first row displays binary masks addressed on the DMD. The second row shows numerically simulated data for the Hartmann sensing step, with centroids marked by crosses, followed by their middle-image profiles in the third row.  (Right panel) The third column, denoted as the maximum-likelihood (ML) reconstruction step, shows a two-aperture mask and simulated data for a partially coherent signal.}
\label{Protocol}
\end{figure}

To compute $\psi_{j}^{\prime} (\xi)$, we employ standard methods from Fourier optics~\cite{Goodman:2004aa}. Assuming the $j$th mode is incident on the DMD, the outcoming signal can be written as 
\begin{equation}
[\mathrm{rect}(x/D) \otimes \mathrm{comb}(x/\Delta)] \psi_j(x) \, ,
\end{equation} 
where $\otimes$ indicates the convolution, and  $D$ and $\Delta$ are the micromirror width and DMD mirror spacing, respectively.  The function $\mathrm{comb}(z)= \sum_{n \in \mathbb{Z}} \delta (z -n )$ accounts for the grating structure of the DMD. This outgoing signal is propagated towards the detector, which spatially separates individual diffraction orders. For our analysis, we consider only one diffraction order.  By applying the Fresnel diffraction integral, the complex amplitude at $z$, the detector plane position, is then obtained as 
\begin{equation}
\label{Sincs2}
\psi_{j}^{\prime} (\xi ) \simeq e^{ik_j x_{0} \sin \alpha} e^{i k_j(\xi - x_{0})^{2}/2z} \sinc \left [   \frac{a}{\lambda z}  ( \xi-x_0 + z\sin \alpha ) \right ] \, ,
\end{equation}
where $\lambda$ is the wavelength. Note the argument of the sinc function [$\sinc x = (\sin x)/x$] takes the form $(\xi-x_0 +z\sin{\alpha})$, where $\alpha$ is the wave deflection angle, indicating that the position of peak intensity depends on the local slope of the wavefront incident on the Hartmann aperture.

Finally, the detected intensity $I(\xi)$ at $\xi$ is determined by the Born rule  $I(\xi)=\Tr(\varrho \, {\Pi_{\xi}})$~\cite{Peres:2002oz}. This reflects the fundamental principle that the measured intensity results from the interplay between the signal state, represented by $\varrho$, and the measurement process, encapsulated by the POVM element $\Pi_{\xi}$.

\begin{figure}[t]
\centerline{\includegraphics[width=0.95\columnwidth]{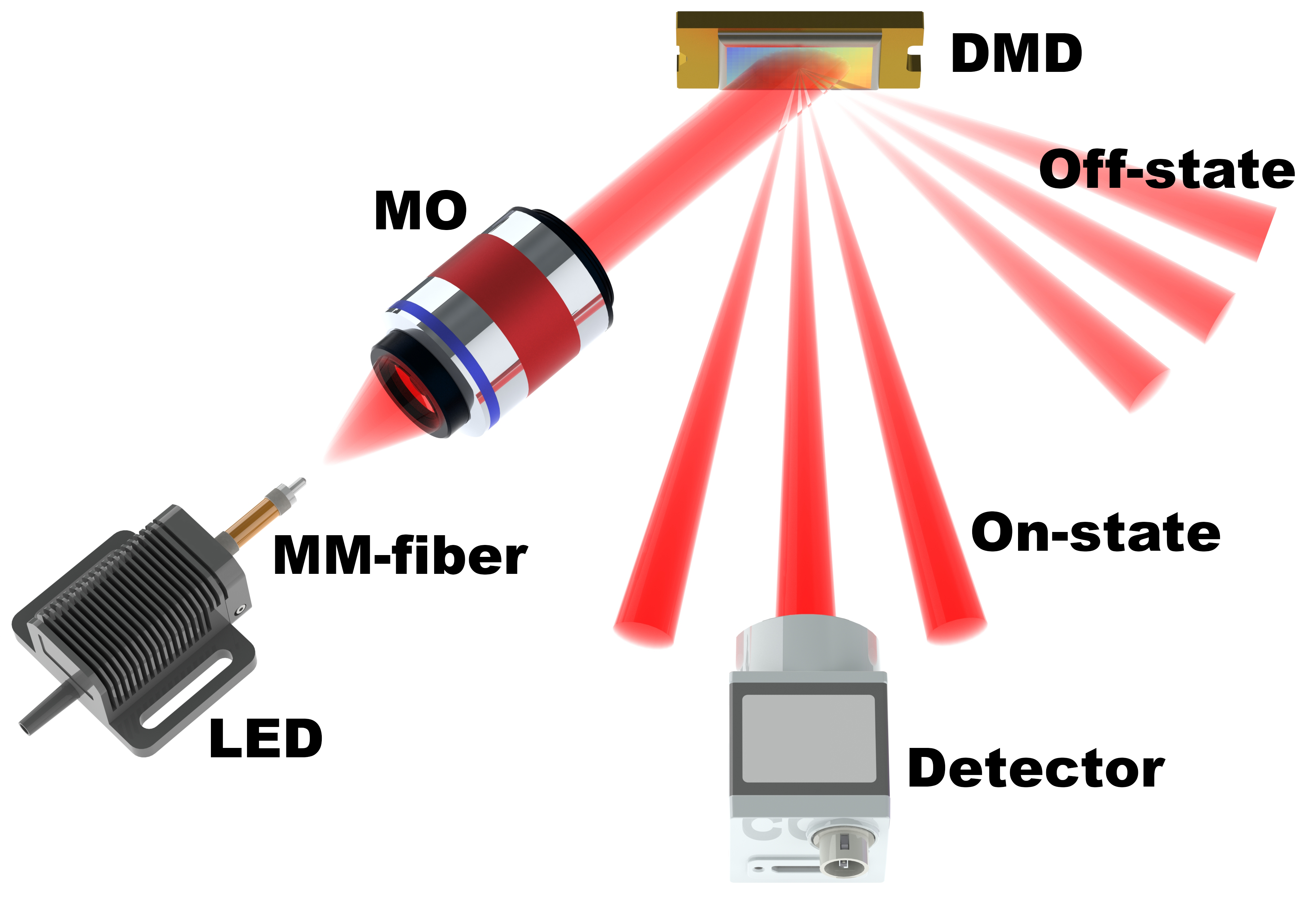}}
\caption{Layout of the Hartmann sensor with the LED-based partially coherent light source used in the experiment. MO stands for microscope objective.}
\label{Setup}
\end{figure}

\section{Reconstruction protocol}
\label{sec:Rec}

We consider a simple scenario where the incoming signal impinges on a two-aperture Hartmann screen, propagating as a statistical mixture of two coherent modes. Under this setup, both the mutual intensity and the measurement reduce to dimension $d=2$.

The first step of our scheme employs a centroid algorithm to reconstruct the $\mathbf{k}$-vector corresponding to individual mask apertures. To this end, we keep only one DMD aperture open, as depicted in Fig.~\ref{Protocol}, ensuring that only a single spot is registered. Among various algorithms proposed for locating spot centroids~\cite{Thomas:2004aa,Wei:2020aa}, the center of gravity (COG) algorithm remains the most widely used due to its simplicity and low computational demand, making it well-suited for our measurement needs. Furthermore, several advanced COG algorithms have been developed to mitigate noise~\cite{Gao:2024aa}. In our implementation, we minimize the root mean squared error between the measured data and our model, which is dependent on the precise spot position. 

In the second step, the information about the $\mathbf{k}$-vectors is utilized to construct the computational basis and the corresponding POVM. 

From the registered intensity $I(\xi)=\Tr(\varrho \, {\Pi_{\xi}})$ and the previous expansions, we can invert this equation and reconstruct $\varrho$ and, consequently, $J (\mathbf{x},\mathbf{x}^{\prime})$. To achieve this, we utilize maximum likelihood  estimation, which allows us to incorporate the positivity constraint on the density matrix. This is essential for both density matrices describing quantum systems and mutual intensity matrices representing classical signals. Given this context, employing a quantum reconstruction algorithm is not merely a peculiarity; it is well justified. We implemented the algorithm described in Ref.~\cite{Hradil:2006aa}, which is specifically designed for biased tomography schemes and operates through an iterative process. 

To increase the accuracy of the proposed reconstruction scheme, the maximum likelihood algorithm was integrated into the minimization procedure, where the centroids of individual spots serve as input variables. We minimize the root mean square error using  the centroid values obtained from the COG algorithm as initial values for the iterative procedure.

Finally, we emphasize that maximum likelihood estimation has been applied to raw Hartmann image data, enhancing accuracy but at the expense of significantly higher computational complexity~\cite{Cannon:1995aa,Dam:2000aa,Barrett:2007aa}. In contrast, our quantum-inspired protocol offers a simple, compact implementation with minimal computational overhead.

\section{Experimental results}
\label{sec:exp}

To validate the method, we use signals with predefined coherence properties. For signal preparation, light from a  light-emitting diode (LED) with a central wavelength of 625~nm  (M625F2, Thorlabs)  is coupled into multimode fibers with  different core diameters. The output from these fibers is then collimated using a microscope objective (TL2X-SAP, Thorlabs) with an effective focal length of $f = 100$~mm.

Given the LED spectral bandwidth, we can consider light
to be quasi-monochromatic. Since the LED source is spatially incoherent, the fiber end-face can be modeled as an incoherent circular planar source. Applying the van Cittert-Zernike theorem~\cite{wolf:2007}, the amplitude of the mutual intensity  $\lvert J (x, x^{\prime})\rvert$  at the collimator output can be expressed as  
\begin{equation}
\displaystyle
| J(x,x^{\prime}) | = 2\lvert u(x) \rvert \, \lvert u(x^{\prime})\rvert \;  \left \lvert \frac{J_1\left ( \frac{\pi w |x - x^{\prime}|}{\lambda f} \right )}{\frac{\pi w |x - x^{\prime}|}{\lambda f}} \right \rvert \, , 
\end{equation}
where $J_1$ is the Bessel function of the first kind and $w$ is the diameter of the fiber core. By using fibers with different core diameters, one can obtain light with varying degrees of coherence. Our reconstruction algorithm provides information about $J (x,x^{\prime})$ in the form of sampled  mutual intensity matrices.

\begin{figure}[t]
\centerline{\includegraphics[width=\columnwidth]{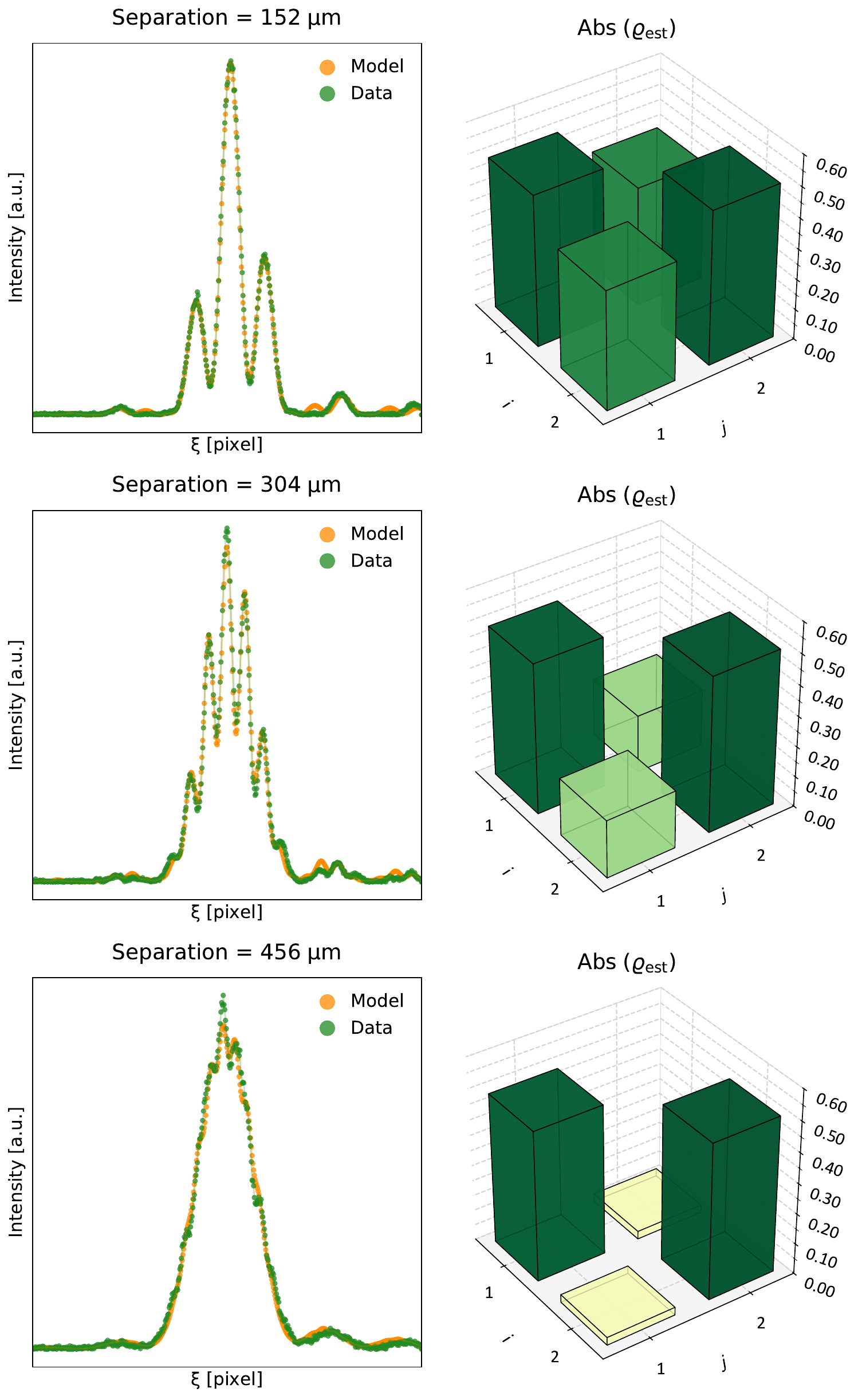}}
\caption{Example of measured data and reconstructed mutual intensity matrices with $200~\mu $m fiber and three different separations of the apertures used to scan the measured field.}
\label{data}
\end{figure}

Our experimental setup is sketched in Fig.~\ref{Setup}. To measure the  prepared signals, we use a DMD (V-6501, Vialux) configured as a reflective binary Hartmann mask. The DMD is positioned such that each micromirror diagonal is perpendicular to the optical table, ensuring that both the incident and reflected signals are in a plane parallel to the table. 

To minimize chromatic blur caused by the diffractive nature of the DMD, the incoming light is filtered using a narrowband filter (FLH633-1, Thorlabs) with a full width at half maximum (FWHM) of 1~nm.  The collimated beam hits the DMD at an angle of $24^\circ$ relative to the normal of the array. This configuration ensures that mirrors tilted at $12^\circ$ reflect light in a direction perpendicular to the chip surface. The reflected light is then detected by an EMCCD camera (Hamamatsu ImagEM X2 C9100-23B), positioned 180~mm from the DMD, allowing for complete overlapping of the signals.

\begin{figure*}[t]
\centerline{\includegraphics[width=2\columnwidth]{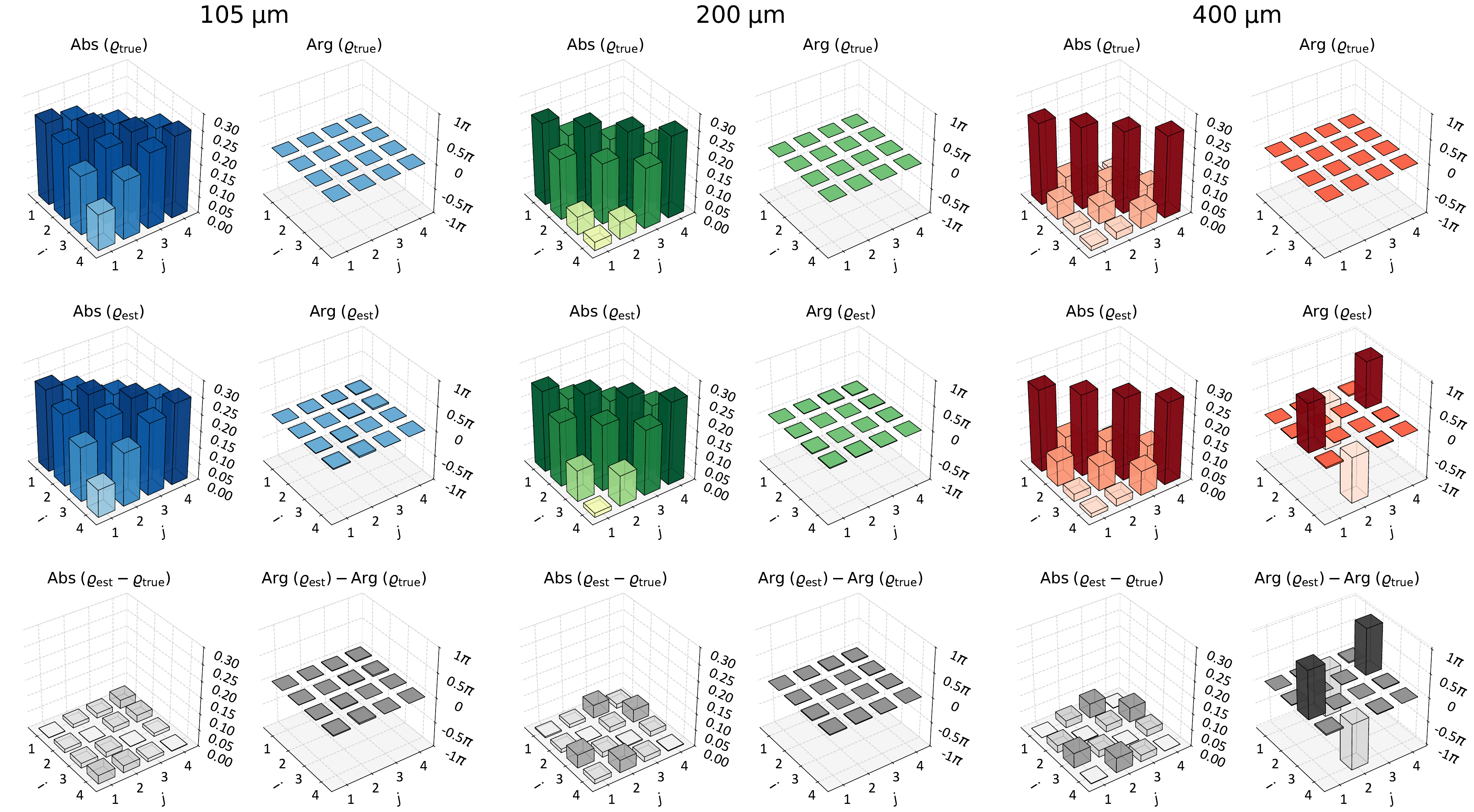}}
\caption{Results for fibers with $105$, $200$, and $400~\mu$m core diameter. The upper row shows the absolute value and the phase of the true $\varrho$, whereas the middle row presents the reconstructed values according to our protocol. The lower row displays the corresponding errors, demonstrating an excellent match with the theoretical predictions.}
\label{results}
\end{figure*}

Once the measured beam is scanned with a two-subaperture mask, we stitch together partial measurements for each combination of measurement points~\cite{Idir:2014aa,Wang:2023aa}. In principle, one might reconstruct the mutual intensity matrix at $N\times N$ points; such a measurement requires $\frac{1}{2}(N^{2}-N)$ samples. This implies a polynomial complexity, but can be easily handled for reasonable values of $N$.  

Our objective is to examine the statistical properties of the prepared signal  as the separation between the Hartmann mask apertures increases. Figure~\ref{data} presents an example from our collected data sets.  The data correspond to a $200~\mu$m fiber core, with two square apertures addressed to the DMD chip  separated by $152$, $304$, and $456~\mu$m, respectively (which corresponds to 20, 40, and 60 pixels, respectively).

As predicted by the van Cittert-Zernike theorem, the degree of coherence decreases as the separation between points increases. This behavior is evident in the off-diagonal elements of the reconstructed matrices, shown in the right column of Fig.~\ref{data}. Additionally, the nearly equal diagonal elements indicate that the intensity of both apertures are roughly identical. To assess the accuracy of our reconstruction, we perform inverse engineering based on the reconstructed data, ensuring the fidelity of our method.

In Fig.~\ref{results} we summarize our results for three fibers with core diameters of $105$, $200$, and $400~\mu$m, respectively. As a proof of principle, we have chosen a  stitching with $N=4$. However, this number can be readily increased, with the only limitations being the spatial size and pixelization of the DMD.  

We present both the theoretical mutual intensity values, derived from the van Cittert-Zernike theorem, and the reconstructed values, along with their associated errors. The absolute value and phase of the mutual intensity are displayed separately. As expected, increasing the fiber diameter leads to a decrease in coherence between individual points, an effect that becomes more pronounced as the separation between points grows.  For the $400~\mu$m fiber,  coherence is confined to a small neighborhood, as indicated by nonzero values appearing only along the diagonal elements.   

Furthermore, the imaginary part of the mutual intensity remains identically zero, as the  matrix is reconstructed in a basis where the propagation directions align with those of the signals. As evident from the plots, the reconstructed values closely match the theoretical predictions, demonstrating the accuracy of our method.

\begin{figure}[t]
\centerline{\includegraphics[width= \columnwidth]{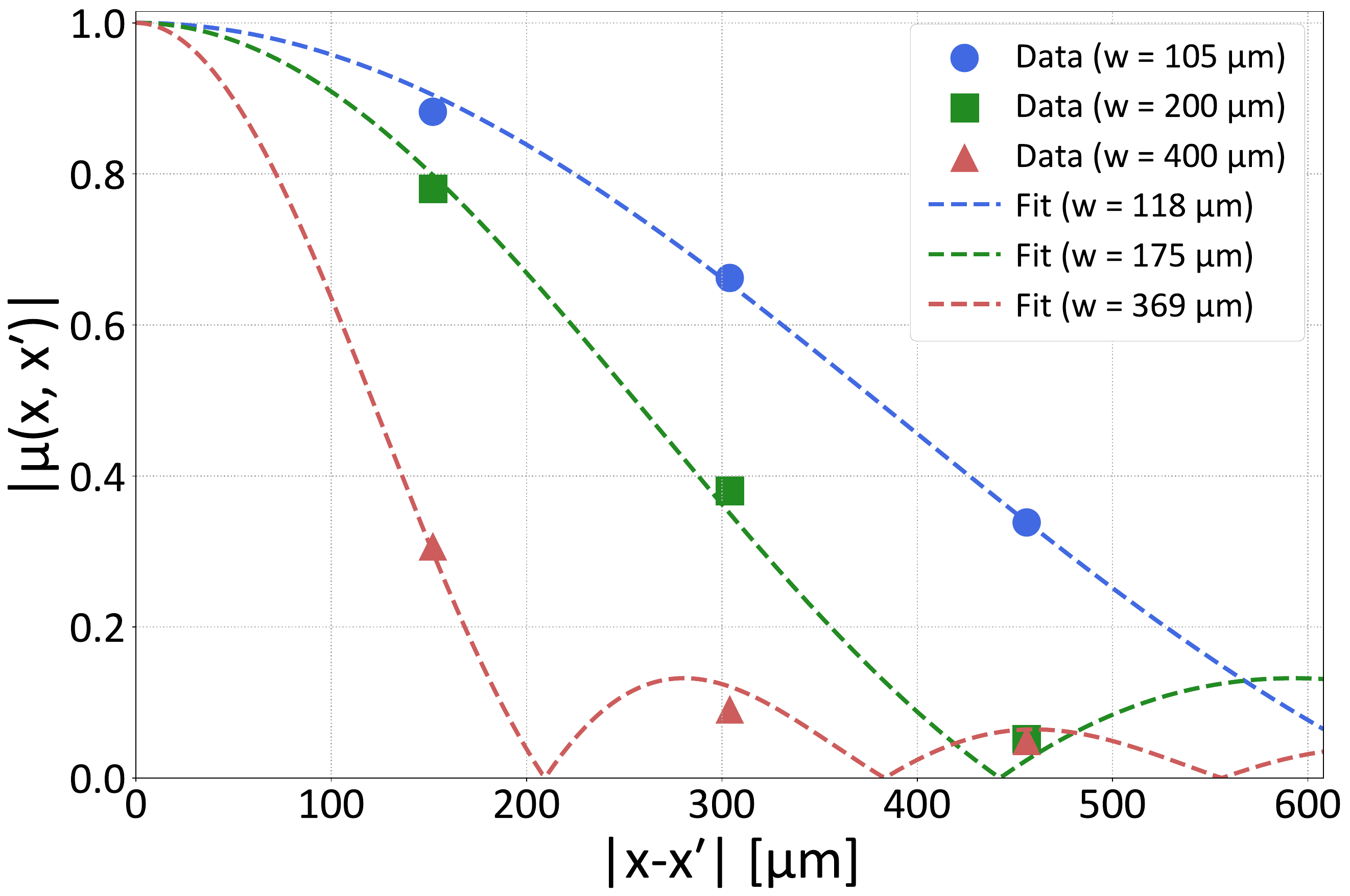}}
\caption{Points represent the measured values of the complex degree of coherence $|\mu(x,x^{\prime})|$ as derived from the data in Fig.~\ref{results}. The curves are the fits to the theoretical predictions from the van Cittert-Zernike theorem \eqref{eq:vZM}.}
\label{fitting}
\end{figure}

To experimentally validate our method, we compare the measured values of the mutual intensity with those predicted by the van Cittert–Zernike theorem. To this end, we normalize $J(x, x^{\prime})$ to obtain the complex degree of coherence~\cite{Mandel:1995qy}, denoted as $\mu(x,x^{\prime})$. For an incoherent circular planar source case, this yields
\begin{equation}
\displaystyle
| \mu (x,x^{\prime}) | = 2    \left \lvert \frac{J_1\left ( \frac{\pi w |x - x^{\prime}|}{\lambda f} \right )}{\frac{\pi w |x - x^{\prime}|}{\lambda f}} \right \rvert \, . 
\label{eq:vZM}
\end{equation}
We fit our experimental results to this function to estimate the values of the diameters $w$ for the three fibers, as shown in Fig.~\ref{fitting}, obtaining  $118~\mu$m, $185~\mu$m, and $369~\mu$m, respectively. Given the limited number of sampled  points, these measurements show satisfactory agreement with the nominal values and serve as a proof of principle for our method. The observed deviations may arise from factors such as partial coherence or finite-aperture effects.

\section{Concluding remarks}
\label{sec:conc}

We have experimentally measured the spatial coherence  of a partially coherent signal, introducing a key innovation: operating a DMD-based Hartmann wavefront sensor in the regime of overlapping spots. This approach overcomes the limitations of traditional wavefront sensors, enabling us to access information beyond the capabilities of conventional reconstruction methods. When the interfering signal arises from spatial points, our method directly yields the $x$-representation of the mutual intensity. 

Whether for improving image quality, enhancing wavefront correction, or understanding quantum phenomena, spatial coherence is a key parameter in modern optics. Our quantum-inspired method paves the way for new possibilities that could be relevant for the field.

\section*{Acknowledgments}
This work has received financial support from the Quant\-Era program (project ApresSF), from Palack\'y University (Grant IGA PRF $2025_{\mbox{--}}005$) and from the Spanish Agencia Estatal de Investigaci\'on (Grant PID2021-127781NB-I00). L.~L.~S.~S. was supported in part by the grant NSF PHY-1748958 to the Kavli Institute for Theoretical Physics (KITP).


\providecommand{\noopsort}[1]{}\providecommand{\singleletter}[1]{#1}%

\end{document}